\documentclass[aps,prd,twocolumn,groupedaddress,nofootinbib]{revtex4-1}

\usepackage{graphics}
\usepackage{amssymb}
\usepackage{amsfonts}
\usepackage{amsbsy}
\usepackage{amsmath}
\usepackage[vcentermath]{youngtab}
\usepackage{graphicx}
\usepackage{amscd}
\usepackage{epstopdf}
\usepackage{amsthm}
\usepackage{enumerate}

\newcommand{\cD}{\mathcal{D}}

\newcommand{\cF}{\mathcal{F}}

\newcommand{\cL}{\mathcal{L}}

\newcommand{\cN}{\mathcal{N}}
\newcommand{\cO}{\mathcal{O}}

\newcommand{\cQ}{\mathcal{Q}}

\newcommand{\cV}{\mathcal{V}}

\newcommand{\bR}{\mathbb{R}}
\newcommand{\bZ}{\mathbb{Z}}

\def\ov{\over}

\def\ket#1{|#1\rangle}

\def\eq#1{(\ref{#1})}
\def\p{\partial}

\def\bp{{\bar{\partial}}}

\def\tW{\widetilde{\mathcal{W}}}

\def\DBPS{\Delta_{BPS}}
\def\bU{{\bar U}}
\def\bV{{\bar V}}

\newcommand{\be}{\begin{equation}}
\newcommand{\ee}{\end{equation}}
\newcommand{\bea}{\begin{equation} \begin{aligned}}
\newcommand{\eea}{\end{aligned} \end{equation}}

\newcommand{\bln}{\begin{align}}
\newcommand{\eln}{\end{align}}
\newcommand{\bst}{\begin{split}}
\newcommand{\est}{\end{split}}
\newcommand{\bi}{\begin{itemize}}
\newcommand{\ei}{\end{itemize}}
\newcommand{\ben}{\begin{enumerate}}
\newcommand{\een}{\end{enumerate}}

\begin{document}


\title{Topological charges in 2d $\mathcal{N}=(2,2)$ theories
and massive BPS states}


\author{Daniel S. Park}
\email[]{dpark {\tt at} scgp.stonybrook.edu}
\affiliation{Simons Center for Geometry and Physics\\
Stony Brook University, \\
Stony Brook, NY 11794-3636, USA}


\date{\today}

\begin{abstract}
We study how charges of global symmetries that are manifest in the ultra-violet definition of a theory are realized as topological charges in its infra-red effective theory for two-dimensional theories with $\mathcal{N}=(2,2)$ supersymmetry. We focus on the charges that the states living on $S^1$ carry. The central charge---or BPS masses---of the supersymmetry algebra play a crucial role in making this correspondence precise. We study two examples: $U(1)$ gauge theories with chiral matter, and world-volume theories of ``dynamical surface operators" of 4d $\mathcal{N}=2$ gauge theories. In the former example, we show that the flavor charges of the theory are realized as topological winding numbers in the effective theory on the Coulomb branch. In the latter, we show that there is a one-to-one correspondence between topological charges of the effective theory of the dynamical surface operator and the electric, magnetic, and flavor charges of the 4d gauge theory. We also examine the topologically charged massive BPS states on $S^1$ and discover that the massive BPS spectrum is sensitive to the radius of the circle in the simplest theory---the free theory of a periodic twisted chiral field. We clarify this behavior by showing that the massive BPS spectrum on $S^1$, unlike the BPS ground states, cannot be identified as elements of a cohomology.
\end{abstract}

\pacs{}

\maketitle

\section{Introduction} \label{s:introduction}

Let us consider a quantum field theory defined at some ultra-violet (UV) scale with a set of massive fields of mass $\sim \mu$, which we schematically denote $\{ \phi_i \}$. In the infra-red (IR) effective theory, describing the physics far below the mass scale $\mu$, the fields $\{ \phi_i \}$ are integrated out and are absent from the path-integral description of the IR field theory, if there is one. Given that the UV Lagrangian had some global symmetry, a natural question to ask is how to detect the symmetries of the UV theory in the IR effective theory. There are well understood classic examples where the charge of a global symmetry of the UV theory manifests itself as a topological charge in the IR \cite{Skyrme, FinkelsteinRubinstein, Faddeev, Coleman, JackiwRebbi, HasenfratztHooft, GoldstoneWilczek, WittenBaryons}. In this paper, we study some two-dimensional examples of theories with $\cN=(2,2)$ supersymmetry in which the correspondence between the charges of the global symmetries in the UV and the topological charges in the IR can be made precise.

The crucial fact that enables us to match the UV and IR charges, is that the BPS central charge of a state in a 2d $\cN=(2,2)$ theory is determined by its topological charges \cite{WittenOlive}. We focus on examples with theories with one twisted chiral multiplet, but our results are expected to generalize to other cases. These topological charges arise when the twisted superpotential of the theory is multi-valued with respect to the value of the scalar $\sigma$ in the twisted chiral multiplet \cite{HLS,BSV}. The multi-valuedness of the superpotential gives rise to a non-trivial topological structure of the space ``$X$" the twisted chiral field $\sigma$ takes values in---in particular, the homology group $H_1(X)$ can be non-trivial. In this case, the states living on the spatial slice $S^1$, or ``$S^1$ states," can carry topological charges valued in $H_1(X)$. The supersymmetry algebra acting on states with a given topological charge has a central term determined by the topological charge \cite{HLS,BSV}. This central charge gives a BPS bound to the mass of states \cite{WittenOlive,FMVW}. When a given theory is an IR effective theory of some UV theory, the topological charges in the IR can be matched with global charges in the UV by comparing the central charge of the supersymmetry algebra of the ``two" theories.

The topologically charged BPS states, the mass of which saturates the lower bound given by a non-zero central charge of the supersymmetry algebra, come in shortened multiplets of the supersymmetry algebra. There is much to be learned about the massive BPS states on $S^1$, the spectrum of which is expected to be more subtle compared to BPS ground states. For example, it has been demonstrated in \cite{BSV} that classical BPS field configurations carrying topological charge do not necessarily survive as BPS states in the quantum theory. In this paper, we show that the massive BPS spectrum depends on the radius of the spatial circle by examining the simplest theory with non-trivial topological sectors---the theory of a free periodic twisted chiral field with a linear twisted superpotential. For generic values of the radius, there are no massive BPS states on $S^1$ in the free theory, while BPS states do appear for discrete values.


In the Morse theory-based approach to BPS spectra \cite{WittenMorse,MooreFlorida,GMWweb,GMW}, the BPS ground states of a sigma model can be identified as the elements of a particular cohomology of the loop space $\cL X$ of the target space $X$.%
\footnote{This approach is also implicit in the work \cite{CVClassification}.}
From this point of view, one might be wary about the claim that the BPS spectra of a theory exhibits unstable behavior under the variation of information contained in the (twisted) superpotential of the theory. We clarify this point by showing that massive BPS states in topological sectors are not elements of a cohomology represented by harmonic forms, but rather distinguished differential forms on loop space that are annihilated by a certain quadratic operator. In the process, we find that the supersymmetry charges can be identified with operators acting on differential forms of $\cL X$. Let us note that the loop space $\cL X$ consists of disjoint components that correspond to different topological sectors of the theory. In topologically non-trivial sectors in which the central charge of the supersymmetry algebra does not vanish, the supersymmetry charges can be constructed by adding a connection to the equivariant differential operators that have been studied, among other places, in \cite{WittenMorse,Teleman,Lillywhite}.%
~\footnote{I would like to thank E. Witten for suggesting this type of modification of the differential.}

We have taken only very preliminary steps in studying massive BPS states on $S^1$ in this paper. It would be interesting to understand what we can learn about them, and to further develop and utilize computational tools \cite{WittenMorse,MooreFlorida,GMWweb,GMW,CFIV,CVIsing,CVClassification,Bershadsky:1993ta} that give access to information about these states.

This paper is organized as follows. In the first two sections, we study examples of theories whose superpotential in the IR is multi-valued in the target space. The first example, which we study in section \ref{s:eg1}, is the Coulomb branch effective theory of a $U(1)$ gauge theory with $n$ charged chiral matter fields. We show that the topological charges of the $S^1$ states of the effective theory is given by $(n-1)$ winding numbers, and identify them with flavor charges in the UV.  In section \ref{s:eg2}, we study what we call the ``dynamical surface operator theory" of 4d $\cN=2$ gauge theories \cite{SW1,SW2,Witten:1997sc,Gaiotto:2009we,Alday:2009aq}. This theory is obtained by taking the surface operator theory of \cite{GGS}, and promoting the complex Fayet-Iliopoulos (FI) parameter to a dynamical twisted chiral multiplet. We show that the topological charges of this theory coincide with the electric, magnetic and flavor charges of the four-dimensional theory. We end by studying the behavior of the BPS spectrum of $S^1$ states in section \ref{s:spectrum}. We first study the spectrum of the theory of a single periodic twisted chiral field with a linear potential in some detail and show that its spectrum varies as the radius of the spatial circle is changed. We then study the problem by recognizing that the states of the theory are differential forms on the loop space of a flat cylinder and characterize the BPS states in this language. In particular, we construct the supersymmetry operators modifying equivariant differential operators by adding a connection, and generalize the construction to arbitrary sigma models with (twisted) superpotentials. Some technical details are collected in the appendices.

\vskip 0.2in
\noindent
{\it Note added:}
{During the completion of this paper, I have become aware that its contents have some overlap with the work \cite{GMW}. I would like to thank the authors of \cite{GMW} for kindly sharing relevant parts of their manuscript before its publication.}

\section{First example: $U(1)$ gauge theory} \label{s:eg1}

Our first and main example is a 2d $\cN=(2,2)$ gauge theory with a $U(1)$ gauge group and $n$ chiral multiplets $\Phi_k$ of charge $q_k$. We turn on generic twisted masses $m_k$ for the chiral fields, and a complex FI term with parameter $t$. Our conventions are such that the imaginary part of the complex FI parameter is the theta angle of the theory. Such theories were studied in great detail in \cite{WittenGLSM,HoriVafa}. The UV theory has a $U(1)^{n-1}$ flavor symmetry, the charges of which we denote by $Q_i$. We denote the gauge charge $Q_0$. The flavor symmetry $U(1)_i$ is defined so that the chiral multiplet $\Phi_k$ has charge $\alpha_{i,k}$. The $\alpha_{i,k}$ are chosen to satisfy $\alpha_{i,k} q_k = 0$, so that they are orthogonal to the gauge charge. Here, let us introduce the unique matrix $\beta_{k,i}$ such that
\be
q_k \beta_{k,i} =0 
\quad \text{and} \quad
\beta_{k,i} \alpha_{i,l}  = \delta_{kl} - {q_l q_k / q^2} \,.
\label{beta}
\ee
While $\alpha_{i,k}$ is an $(n-1) \times n$ matrix, it is useful to think of it as an invertible linear map from the $(n-1)$-dimensional subspace $S_\cF=\{ (x_k) : q_k x_k =0\}$ of $\bR^n$ to $\bR^{n-1}$. $\beta$ is the inverse map of $\alpha$ from $\bR^{n-1}$ to $S_\cF$.

The $U(1)$ gauge multiplet has a complex scalar field $\sigma$ as a component. At some IR scale $\mu_{IR} \ll m_k$, when this complex scalar field takes values such that $(q_k \sigma + m_k) \gg \mu_{IR}$, the chiral fields can be integrated out from the theory. The effective theory in the IR is a theory with a single $U(1)$ vector multiplet with a twisted superpotential \cite{WittenGLSM}:
\be
2 \pi \tW = -t_{eff} \Sigma - \sum_k (q_k \Sigma+m_k) \log (q_k \Sigma+m_k)/\mu_{IR} \,,
\label{tsp}
\ee
where we have dropped a constant shift, which does not affect the physics in anyway. The field $\Sigma$ is the twisted superfield that has $\sigma$ as its lowest component. The charges $Q_i$ are then identified with the topological charges of the effective theory:
\be
Q_i = {1 \ov 2 \pi} \int_{L} dx \, \p_x \sum_k \alpha_{i,k} \arg (q_k \sigma+m_k)
\ee
where the integral is taken over the spatial slice $L$ of the QFT. This equation can be obtained by integrating expectation value of the flavor current component $\alpha_{i,k} j_k^0$. In particular, it can be understood as coming from the term $\bar \psi_k \gamma^0 \psi_k$ in $j_k^0$, due to the fact that $(q_k\sigma+m_k)$ is the effective complex twisted mass of the fermions in the chiral multiplet $\Phi_k$ \cite{GoldstoneWilczek}. Here, $\psi_k$ is the Dirac fermion in the multiplet $\Phi_k$. We prove this claim when $L$ is an $S^1$ with more rigor shortly.

Before doing so, let us note that when $L$ is a circle, $Q_i$ is given by the winding number of the phase of $\prod_i (q_k \sigma+m_k)^{\alpha_{i,k}}$ around $L$, i.e.,
\be
Q_i = {1 \ov 2\pi} \sum_{k} \alpha_{i,k} \Delta(\arg (q_k \sigma + m_k)) \,.
\label{Qi}
\ee
These winding numbers span the complete lattice of topological charges of a state on $S^1$ in the IR theory. This is because the target space of the IR theory, due to the superpotential \eq{tsp}, can be thought of as an $n$-punctured plane $X$ with punctures $P_k$ at $\sigma = -m_k/q_k$. Therefore the lattice of topological charges, or the winding numbers of $S^1$ states live inside $H_1(X,\bZ)$, spanned by the one-cycles $C_k$ surrounding the puncture $P_k$. The topological charges of a state on the compact manifold $S^1$, however, cannot span the entire vector space $H_1(X,\bZ)$. This is because of Gauss' law, which imposes the state to be neutral under the $U(1)$ gauge symmetry. This restricts the winding number $w = w_{k} C_k$ of the state to satisfy
\be
q_k w_k = q_k \Delta(\arg (q_k \sigma + m_k)) =0 \,.
\label{Gauss}
\ee
We see that the winding numbers \eq{Qi} form a basis for this subspace of $H_1(X,\bZ)$, and hence represent all topological charges an $S^1$ state can have.

We can prove \eq{Qi} in two different ways. The first is to use a supersymmetric version of bosonization \cite{RocekVerlinde,HoriVafa}, which is often referred to as ``dualization" in the literature. The chiral multiplets $\Phi_k$ can be dualized to periodic twisted chiral multiplets $Y_k \sim Y_k + 2\pi i$, where
\be
\p_x {\rm Im} y_k = -2 \pi j^0_k \,.
\ee
Here, $x$ is used to denote the spatial direction while $y_k$ is the lowest component of $Y_k$. In particular, this implies that the charge obtained by integrating $j^0_k$ along space is to be identified with the negative of the winding number of $Y_k$. Then it has been shown that the initial theory is dual to a $U(1)$ theory coupled to these twisted multiplets via the twisted superpotential \cite{HoriVafa}:
\be
2 \pi \tW = - t\Sigma + \sum_k (q_k\Sigma + m_k) Y_k + \mu \sum_k e^{-Y_k}  \,.
\label{tspY}
\ee
Just as before, when $(q_k \sigma + m_k) \gg \mu_{IR}$, the twisted chiral fields $Y_k$ can be integrated out to yield an IR effective theory of a single vector multiplet. This leads to the effective superpotential \eq{tsp} as expected. Meanwhile, upon integrating $Y_k$ out using \eq{tspY}, we find that $y_k$ is to be identified with $-\arg (q_k \sigma +m_k)$. Hence, the winding number of $Y_k$ is identified with the negative of the winding number of $\sigma$ around the point $-m_k/q_k$. Since the flavor charge $Q_i$ is given by  the winding number of $-\sum_k \alpha_{i,k} Y_k$, we arrive at \eq{Qi}.

There is another way of getting at the identity \eq{Qi} without relying on any dualities. This involves the central charge of the supersymmetry algebra of the theory. By acting on the chiral fields with the supersymmetry operators, we find that
\be
\{ \bar \cQ_+, \cQ_- \} \Phi_k = 2 (q_k \sigma + m_k ) \Phi_k \,,
\label{chiral}
\ee
where we have used the conventions of \cite{HoriVafa}. Defining the central charge $Z$ of the supersymmetry algebra by
\bea
\{ \cQ_+, \bar \cQ_+ \} &= 2(H-P) \,, &
\{ \cQ_-, \bar \cQ_- \} &= 2(H+P) \,, \\
\{ \bar \cQ_+, \cQ_- \} &\equiv 2iZ \,, &
\{  \cQ_+, \bar \cQ_- \} &= -2i\bar Z \,,
\label{susy}
\eea
the equation \eq{chiral} may be re-written as a relation between the central charge and the gauge / flavor charges:
\be
Z = -i \left( \sigma + {\sum_{k} m_k q_k  \ov q^2 }\right)Q_0 -i \sum_{i,k} m_k \beta_{k,i}  Q_i \,.
\ee
The bar on $Z$ in equation \eq{susy} has been used to denote complex conjugation. Using
\be
Q_0 \Phi_k = q_k \Phi_k, \quad Q_i \Phi_k = \alpha_{i,k} \Phi_k \,,
\ee
and the definition \eq{beta} of $\beta$, we can check that equation \eq{chiral} is indeed satisfied for the chiral fields $\Phi_k$.

For states coming from quantizing the phase space of fields on $S^1$, $Q_0$ is always zero:
\be
Z =- i \sum_{i,k} m_k \beta_{k,i} Q_i \,.
\label{ZUV}
\ee
Note that the central charge $Z$ gives a lower-bound for the mass $m$ of $S^1$ states, i.e., $m \geq |Z|$. BPS states are defined to be states that saturate this bound. A massive BPS multiplet consists of two components, as opposed to the four of a long multiplet \cite{CFIV}.

Meanwhile, in the IR theory, the central charge $Z$ of the supersymmetry algebra can be computed to be \cite{WittenOlive,FMVW,dGIT}
\be
Z = \Delta \tW (\sigma)\,,
\label{top charge}
\ee
where $\Delta \tW (\sigma)$ denotes the integral of $\p_x \tW (\sigma)$ along the spatial slice. Now when $\tW$ is single valued for given values of the twisted chiral fields, the only way $Z$ can be non-zero is when the spatial slice is a line and $\sigma$ interpolates between the saddles of $\tW$ from one end of the line to the other \cite{WittenOlive,FMVW}. In this case, states on $S^1$ cannot have a non-trivial central charge. Meanwhile, when $\tW$ is multivalued with respect to $\sigma$, and the different values can be approached by continuous variations of the fields, $\Delta \tW$ can have non-trivial values even on compact spatial slices, such as $S^1$ \cite{HLS,BSV}. The twisted superpotential of the effective $U(1)$ theory in question is precisely of this nature \cite{WittenGLSM}. In fact, on $S^1$, the central charge is related to the winding numbers around $P_k$ by
\be
Z = \Delta \tW = -{i \ov 2\pi} \sum_k m_k \Delta(\arg (q_k \sigma + m_k))
\label{ZU1}
\ee
due to the superpotential \eq{tsp}. Again, we stress that for $S^1$ states, the winding numbers are restricted by the Gauss constraint \eq{Gauss} which is crucial for arriving at the above equation. Equating equations \eq{ZUV} and \eq{ZU1}, we arrive at the identification \eq{Qi}.

We thus have derived that the flavor charges of the UV gauge theory become topological charges, or winding numbers, of the IR theory. We have also derived the BPS central charge of states in $S^1$ with respect to the topological charges in equation \eq{ZU1}.

\section{Second Example: Worldvolume theory of a dynamical surface operator}\label{s:eg2}

Our next example is that of a ``dynamical surface operator" (DSO) in a 4d $\cN =2$ $SU(N)$ supersymmetric gauge theory \cite{SW1,SW2,Witten:1997sc,Gaiotto:2009we,Alday:2009aq}. Our dynamical surface operator theories come from a slight modification of the class of surface operators \cite{Gukov:2006jk,Gaiotto:2009fs,Gaiotto:2010be} studied in \cite{GGS}. The surface operators of \cite{GGS} are 2d defects of the 4d theory, of which the world-volume degrees of freedom consist of a $U(1)$ vector field and $N$ chiral multiplets, with a Fayet-Iliopoulos term $\tW = t\Sigma$. As before, $\Sigma$ is the field strength multiplet of the vector field. The surface operator theory becomes coupled to the bulk theory by coupling the adjoint scalar field of the 4d vector multiplet as a twisted mass of the $SU(N)$ flavor symmetry of the chiral multiplets in the 2d theory. The dynamical surface operator theory is obtained by promoting the FI parameter $t$ to a twisted chiral superfield, which we denote $T$. While many aspects of the theory of the DSO depend on the details of how the FI parameter is made dynamical, the topological charges and the central charge of the supersymmetry algebra do not depend on these details, but only on the effective twisted superpotential of the theory, which has been computed in \cite{GGS}. Let us study the topological charges of the $S^1$ states of this theory, and relate them to the central charge of the supersymmetry algebra.

The effective dynamical surface operator theory is obtained by moving to a point $u$ in the Coulomb branch of the 4d theory and integrating out all the massive fields of the 4d and 2d theory at that point. As a result of \cite{GGS}, the effective theory of a DSO then becomes a 2d sigma model into the Seiberg-Witten curve $\Sigma_u$ \cite{SW1,SW2} of the 4d theory at $u$, which is given by an $N$-fold cover of the complex $t$-plane. The IR theory has a twisted superpotential \cite{GGS,GMN}:
\be
\tW = \int^p \lambda_{SW,u}
\label{DSOtsp}
\ee
where $p$ is a point on $\Sigma_u$, and $\lambda_{SW,u}$ is the Seiberg-Witten differential at $u$ \cite{SW1,SW2}. Now the superpotential \eq{DSOtsp} is not singled valued with respect to point $p$, which is of course, associated to the topology of the target space. First of all, the Seiberg-Witten curve is a Riemann surface and has non-trivial cycles. Also, the Seiberg-Witten differential $\lambda_{SW,u}$ has a collection of simple poles $P_\mu$ on $\Sigma_u$, given that generic twisted masses $m_F$ have been turned on for the Cartan subgroup of the four-dimensional flavor symmetry. While the number of poles $N_P$ is in general greater than the rank $R_f$ of the flavor symmetry of the four-dimensional theory, the number of independent residues is given precisely by $R_f$. In other words, there are $(N_P-R_f)$ linear combinations of poles $\widetilde P_k$ such that the residues of the Seiberg-Witten differential at
\be
\widetilde P_k = \sum_\mu c_{k,\mu} P_\mu , \quad i=1,\cdots,(N_P-R_f)
\ee
are trivial. Denoting the cycles surrounding $P_\mu$ as $C_\mu$, we find that the linear combination of cycles $\widetilde C_k = \sum_\mu c_{k,\mu} C_\mu$ are invisible to the Seiberg-Witten differential. We hence see that the lattice of topological charges of the $S^1$ states of the theory can be identified with $H_1(\Sigma\setminus\{ P_\mu\},\bZ) / \Lambda $, where $\Lambda$ is a subgroup of the homology group generated by $\widetilde C_k$.

Hence there are $2(N-1)$ topological charges $n_i$ and $m_i$ which are winding numbers around non-trivial cycles of the Seiberg-Witten curve and $R_f$ topological charges $s_F$ coming from the linearly independent combination of cycles $C_F$ that surround the linear combination $P_F$ of poles of $\lambda_{SW,u}$. The topological charges $n_i$, $m_i$ and $s_F$ of the theory can be identified with the electric, magnetic and flavor charges of the 4d theory \cite{SW1,SW2,GGS}. Note that the 4d hypermultiplets have been completely integrated out of the theory, and that the UV flavor symmetry manifests itself in the two-dimensional effective theory as a topological charge, as advertised. Moreover, the central charge $Z$ of the supersymmetry algebra, given that the spatial slice of the 2d theory is $S^1$, is given by
\be
Z = \Delta \tW = \sum_{i=1}^{N-1} (n_i a^i + m^i a_{D,i}) + \sum_{F=1}^{R_f} s_F m_F \,,
\ee
where $a^i (u)$ and $a_{D,i} (u) $ are the electric-magnetic weights, while $m_F$ are the twisted masses for the 4d flavor symmetries \cite{SW1,SW2,GGS}. This follows from the definition of $a^i$, $a_{D,i}$ as integrals of $\lambda_{SW,u}$ over the electric and magnetic cycles of the Riemann surface, and $m_F$ as the residue of the Seiberg-Witten differential at $P_F$. As a result, the charges and the mass formula of the BPS states of this 2d theory on $S^1$ coincide with the charges and the mass formula for the BPS particles of the 4d theory on $\mathbb{R}^3$.

\section{BPS states with topological charges} \label{s:spectrum}

A natural question to ask in light of our results is how to calculate the spectrum of BPS states on $S^1$. The BPS spectrum, however, is expected to be rather non-robust. To be more precise, we expect the BPS spectrum of the two-dimensional theory to depend on the details of the spatial slice. For example, it is clear that the BPS spectrum will depend on the spin structure of the compact space $S^1$. Unless the spin structure is fixed to be Ramond-Ramond (RR), the $\cN=(2,2)$ supersymmetry is broken by boundary conditions. More surprisingly, we can see that the spectrum with RR spin structure jumps as the radius of $S^1$ varies, even for a theory as simple as the free theory with a linear superpotential. We exhibit that the BPS spectra change drastically with respect to the dimensionless central charge $rZ$, where $r$ is the radius of the spatial circle.

This may be a cause of alarm for some, as BPS ground states of a sigma model into $X$ can be understood as elements of a particular cohomology of $\cL X$, the free loop space of $X$ \cite{WittenMorse,MooreFlorida}. We show that this is no longer true for massive BPS states. Given that the theory has non-trivial topological charges, the loop space $\cL X$ decomposes into disconnected components $\cL X|_{\vec Q = \vec w}$, one for each topological sector of charge $\vec w$. A massive BPS state of charge $\vec w$ is a differential form in $\cL X|_{\vec Q = \vec w}$, annihilated by a differential operator $\DBPS$ that we construct shortly. The BPS ground states, which lie in the topological sector $\cL X|_{\vec Q = \vec 0}$, are defined to be differential forms annihilated by the positive semi-definite Laplacian operator constructed from supercharges, which correspond to linear operators $D$ and $D^*$ acting on differential forms in the loop space.%
~\footnote{Here we momentarily convert to describing theories with $\cN=(1,1)$ supersymmetry for simplicity of presentation. The supercharges are identified with modified versions of the exterior derivative $d$ and its dual $d^*$ that act on differential forms of $\cL X$ in this case. In $\cN=(2,2)$ theories, the supercharges are modified versions of Dolbeault operators and their duals, as we present at the end of the section.} Thus the ground states $\Psi$ of the theory must be ``harmonic" in the supercharges:
\be
D \Psi = D^* \Psi = 0\,.
\ee
This is why they represent elements of the cohomology of $\cL X|_{\vec Q = \vec 0}$. Meanwhile, the operator $\DBPS$ that annihilates massive BPS states is not positive semi-definite. While the basis of massive BPS states can be organized such that each element $\Psi_\alpha$ is annihilated by a differential operator $Q_\alpha=e^{i\theta_\alpha} D + e^{-i\theta_\alpha} D^*$, $\Psi_{\alpha}$ is not annihilated by $Q_\alpha^*$. Thus these states are not elements of the $D$-cohomology, and are not guaranteed to be stable under deformations of the theory.


\subsection{Free theory} \label{ss:free}

Let us now consider a theory with one twisted chiral multiplet $Y$ with periodicity $Y \sim Y+2 \pi s i$ and twisted superpotential $\tW = MY$. This theory is a sigma-model into a flat cylinder. We take $s$ to be an arbitrary real number. The topological charge of the states on $S^1$ is given by the winding number $w$ around the imaginary direction of $Y$, and the central charge of the supersymmetry algebra is given by
\be
Z = \Delta \tW = 2 \pi i Ms w  \,.
\ee
Meanwhile, the twisted superpotential merely adds a constant $|M|^2$ to the Lagrangian of the theory. Free theories of periodic superfields are well understood. In particular, at generic values of the radius of $S^1$, the free theory does not have any states of mass $2 \pi|M|s w$ as we see shortly. Furthermore, the theory is devoid of any classical BPS winding configurations. When the radius $r$ of the $S^1$ satisfies
\be
|M| r = w s \quad
\label{r}
\ee
for some $w \in \bZ$, however, there exist classical time-independent BPS configurations with winding number $\pm w$ of which the mass saturates the BPS bound.

Let us examine the quantum $S^1$ states of this theory with RR spin structure. The spectrum of the quantum theory is well known, and is explained in great detail in standard texts \cite{Polchinski1,Polchinski2}. The states are constructed from the ``ground states" $\ket{p;n,w;i,j}$ that are labeled by three numbers---the momentum $p$ in the non-compact direction, the discrete momentum $n$ and the winding number $w$ in the compact direction---and the spinor indices $i,j$. These states are in the $\bf{2} \times \bf{2}$ representation of $Spin(2)$. All states of the quantum theory can be obtained by acting on the ground states in a given momentum-winding sector with creation operators of massive modes on $S^1$. The mass of a state in a momentum-winding sector labeled by $(n,w)$ then satisfies
\be
m^2  \geq \pi^2 r^2 \left( {n^2 \ov s^2 r^2} - {w^2 s^2 \ov r^2} + |M|^2 \right)^2 + 4 \pi^2 w^2 s^2 |M|^2  \,,
\ee
where the inequality is saturated if and only if $p=0$ and the state is a ground state in the $(n,w)$ sector. What this equation shows is that for fixed winding number $w$, the BPS bound is saturated if and only if
\be
|M|r = \sqrt{ {w^2 s^2 } - {n^2 \ov s^2} }
\label{qr}
\ee
for some integer $n$. Now the momentum of this state is given by $P=2 \pi nw/r$, hence we see that the condition for there to be a time-independent BPS state ($n=0$) is precisely given by the classical formula \eq{r}. While topologically charged BPS field configurations are not, in general, guaranteed to survive as BPS states in the quantum theory \cite{BSV}, in the simple case of the free theory, we see they do.

Let us note that nothing special happens at the radius \eq{qr}---no new states are introduced to the spectrum, or taken away. At this radius, the four states $\ket{0;n,w;i,j}$, with fixed $n$ and $w$, happen to have mass $2 \pi |M s w|$ and split into two BPS multiplets, whereas at a generic radius, these states form one long representation of the supersymmetry algebra. It is clear that for the free theory, the spectrum of BPS states change with $r$, and in fact, for generic $r$, there are no BPS states at all.

From this example, we see that massive BPS states on $S^1$ can appear and disappear as the radius of the spatial slice is tuned, and that there may be no BPS states at certain, or even most radii. This does not violate any index theorem, as the BPS states appear and disappear in pairs. In fact, this is something expected, as the Cecotti-Fendley-Intriligator-Vafa (CFIV) index ${\rm Tr} F (-1)^F$ \cite{CFIV}, which is specialized to counting massive BPS states, vanishes on the torus due to CPT invariance \cite{CVIsing}. This implies that $S^1$ states with topological charge, even if they exist, come in pairs and cancel out in the index, just as in the free theory.

Such behavior of the states can be attributed to the fact that the massive BPS states cannot be identified as elements of a cohomology, unlike the zero-energy ground states of the theory. This can be explicitly demonstrated for the free theory studied above. All the salient features can be exhibited, in fact, with a $\cN=(1,1)$ theory obtained by only retaining the compact scalar field $\phi \sim \phi + 2\pi s$ and a single Majorana fermion \cite{WittenOlive}. The linear superpotential $\cV = -i M \phi$ of the theory is now real, i.e., $M$ is a real mass. This theory is an $\cN=(1,1)$ supersymmetric sigma model whose target space is a circle.

The states of these theories are constructed from ground states
\be
\ket{n,w}, ~
\psi_0 \ket{n,w}
\ee
that are labelled by the momentum and the winding number as before, except now the RR-vacuum has degeneracy-two. We have made the presence of the fermion zero mode $\psi_0$ explicit. Note that $\psi_0^2=1$. The states are built by acting on the vacua by real oscillators $\alpha_{-l}$ and $\psi_{-l}$:
\be
\left(\prod_{l=1}^\infty \alpha_{-l}^{n_l} \psi_{-l}^{f_l} \right) \psi_0^{f_0} \ket{n,w} \,.
\label{states}
\ee
The powers of the fermion oscillator $f_l$ for $l>0$ are restricted to be less than two, for the state to not vanish.

These states, which form a basis for the wave function of the theory, can be understood as complex-valued differential forms on the free loop space $\cL S^1$ of the circle, which is just the space of all the maps from the spatial circle of the quantum field theory to the target space. We thus use the term ``states" and ``differential forms" interchangeably. Any such map $\phi$ can be decomposed into Fourier components:
\be
\phi(x) = \phi_0 + {ws \ov r} x + \sum_{l=1}^\infty \phi_{l} \sin {2 \pi l x \ov r} \,.
\ee
Now the free loop space decomposes into discrete components $\cL S^1_w$ labelled by the winding number $w$:
\be
\cL S^1_w = S^1 \times \bR_1 \times \bR_2 \times \cdots \,.
\ee
The coefficients $\phi_i$ are the coordinates on this infinite dimensional space. In particular, $\phi_0$ spans the target circle. A particular basis of differential forms on this space consists of the forms
\be
e^{2\pi i n \phi_0 /s} d \phi_0^{f_0}
\wedge \left( \bigwedge_{l=1}^\infty \Psi_{l,n_l} (\phi_l)  d\phi_l^{f_l} \right) \,.
\label{forms}
\ee
Here, $\Psi_{l,n_l}$ is the level-$n_l$ eigenfunction of the harmonic oscillator of mass $l$. The differential forms \eq{forms} precisely correspond to the eigenstates \eq{states} of the free theory.

Let us now understand how the massive BPS states of the theory are characterized. By examining the supersymmetry operator, we see that the BPS states should be in the ground state of the non-compact directions. The only relevant geometry is the circle $S^1$ of radius $s$ in $\cL S^1_w$. We now focus our attention on the component of the wave function of the theory along the $S^1$ direction, i.e., differential forms on $S^1$.

The real supersymmetry generators $\cQ_1$ and $\cQ_2$ satisfy the algebra \cite{WittenMorse}
\be
\cQ_1^2 = H-P, \quad
\cQ_2^2 = H+P, \quad
\{ \cQ_1 \c,\cQ_2 \} =4 \pi ws M
\label{SUSY11}
\ee
in the sector with winding number $w$. The supersymmetry operators are differential operators on the forms in the loop space. Restricting to the differential forms along the $S^1$, the operators can be identified. Writing \cite{WittenMorse}
\be
{\cQ_1 \ov \sqrt{2\pi}} = i^{1/2} D + i^{-1/2} D^*,~
{\cQ_2 \ov \sqrt{2\pi}} = i^{-1/2} D + i^{1/2} D^* \,,
\label{QD}
\ee
we find that
\be
D = d + ws \iota_{\phi_0} + M d\phi_0 \wedge, ~
D^* = d^* + ws d \phi_0 \wedge + M \iota_{\phi_0} \,,
\label{Ddef}
\ee
where $d$ is the exterior derivative acting on differential forms. Here we have set $r=1$ and absorbed it into $M$, for notational simplicity. This is a slight generalization to an equivariant differential of \cite{WittenMorse}---the ``connection" $M d\phi_0$ has been further added to the equivariant differential with respect to the Killing vector field along the circle. The star superscript acting on an operator $\cO$ denotes its dual defined by
\be
\int \cO^* \alpha \wedge * \beta \equiv
\int \alpha \wedge * \cO \beta \,,
\ee
where the star operator is used to denote the operation
\be
(* \eta)_{i_1 \cdots i_{n-k}}
= {1 \ov k! } \sqrt{\det g} \,
\eta^{j_1 \cdots j_k}
\epsilon_{j_1 \cdots j_k i_1 \cdots i_{n-k}}
\ee
for differential forms on an $n$-manifold.%
~\footnote{While the star operator of this equation can be identified with the standard definition of the Hodge star operator \cite{GriffithsHarris} for real differential forms, the action of the two operators on complex differential forms differ by a complex conjugation of the components.}
In particular, for the differential operator $d$, $d^* = * d *$ in one dimension. The operator $\iota_{\phi_0}$ denotes the interior product with the Killing vector $\p_{\phi_0}$. It is simple to check the following operator relations:
\bea
D^* D +DD^*=& \,\, d^*d + dd^*
+ M^2 +w^2s^2  \,, \\
D^2 = ws \cL_{\phi_0} + Mws \,, & \quad
(D^*)^2 = -ws \cL_{\phi_0} + Mws \,.
\eea
$\cL_{\phi_0}$ is the Lie derivative along $\p_{\phi_0}$. These operators, along with $(DD^*-D^*D)$, mutually commute. An easy way to check this is to use the fact that $D^2+(D^*)^2$ is central. It follows that both $D$ and $D^*$ commute with $D^2$ and $(D^*)^2$. This is enough to verify the claim above.

Now the Hamiltonian and the momentum operator are given by
\bea
H &= 2\pi( D^* D + D D^*) \\
P &= 2 \pi i \big\{ (D^*)^2 -D^2 \big\} = - 2\pi i ws \cL_{\phi_0} \,.
\eea
The zero-energy states of this Hamiltonian are elements of the cohomology defined by the operator $D$. Since $d^* d + dd^*$ is a positive semi-definite operator, $H$ cannot be zero unless $M=w=0$. Therefore the $D$-cohomology is empty in topologically charged sectors. On the other hand when $M=w=0$, the $D$-cohomology is nontrivial---it is just the ordinary cohomology of the circle, and has two elements. These two elements are the ground states of the theory with $M=0$.

The massive BPS states cannot be elements of the empty $D$-cohomology. Rather, they are defined as forms annihilated by the operator:
\be
\Delta_{BPS} = DD^* - D^*D = -{i \ov 4\pi} \left( \cQ_1 \cQ_2 - \cQ_2 \cQ_1\right)\,.
\label{BPS}
\ee
This is due to the relation
\be
4 \pi^2 \Delta_{BPS}^2 = H^2 - P^2  - (2 \pi wsM)^2 \,,
\label{BPSHP}
\ee
i.e., the BPS bound is saturated for the mass of states with $\Delta_{BPS}=0$. Note that $H$, $P$ and $\Delta_{BPS}$ mutually commute and are Hermitian---the BPS states are hence taken to be eigenstates of $P$ as well as of $H$. A consequence of \eq{BPSHP} is that a simultaneous eigenstate of $H$ and $P$ automatically is also an eigenstate of $\Delta_{BPS}^2$.

The BPS states can also be recognized in a more standard way, i.e., as states being annihilated by a linear combination of supercharges. Given an eigenstate $\Psi_{h,p}$ of $H$ and $P$,
\be
\DBPS \Psi_{h,p} = 0
\ee
if and only if
\be
(e^{i\theta}D + e^{-i\theta} D^*) \Psi_{h,p} = 0 
\ee
for $\theta$ such that
\be
e^{2i\theta} = {2 \pi wsM+ ip \ov h} \,.
\ee
While the proof is standard, we have added it in appendix \ref{ap:a} for sake of completeness.

As noted, the eigenstates on the circle are given by the differential forms
\be
e^{2\pi i n \phi_0 /s},  \quad e^{2\pi i n \phi_0 /s} d \phi_0 \,,
\ee
for integer $n$. From \eq{BPSHP}, we find that a BPS state of the free theory at hand must satisfy
\be
{n^2 \over s^2} -{w^2 s^2} + M^2 = 0
\ee
for integers $n$ and $w$, as we have obtained before in the Hamiltonian picture. We thus recover the result that BPS states can only exist for values of $|M|r$ that satisfy the condition \eq{qr}.

\subsection{General case}\label{ss:general}

What we have said so far can be applied, with some modification, to topologically massive BPS states of sigma models with $\cN=(1,1)$ supersymmetry in general. $\cL X$, the free loop space of $X$ decomposes into components $\cL X |_{\vec Q = \vec w}$, where $\vec w$ labels the topological charges, or equivalently, winding numbers. The supersymmetry operators can be constructed out of differential operators on $\cL X |_{\vec Q = \vec w}$. To be more precise, the relations \eq{QD} hold for the differential operators $D$ in the free loop space of the target manifold:
\be
D \equiv d + \iota_{v_K} + A^u\wedge \,.
\ee
$v_K$ is the Killing vector inherited from the isometry along the spatial circle of the 2d theory. $A^{u}$ is the dual one-form to a vector field $u$, i.e.,
\be
A^{u} \wedge = \iota_{u}^* \,.
\ee
These operators imply the supersymmetry algebra with central charge $Z$ \eq{SUSY11}, given that $A^u$ satisfies
\be\label{Au properties}
\cL_{v_K} A^u =0, \quad
d A^u = 0\,.
\ee
The vector field $v_K$ and the relevant one-form $A^u$ can be explicitly written for the cases of interest, i.e., in topologically non-trivial components of the loop space $\cL X$, and the properties \eq{Au properties} can be explicitly verified. The details of this computation are presented in appendix \ref{ap:diff loop}. By definition of the Lie derivative, it follows that the pairing
\be
\iota_{v_K} A^u = Z
\ee
is constant. In topologically nontrivial sectors with $Z \neq 0$, the cohomology with respect to $D$ is trivial since $\{ D, D^* \} \geq |Z| > 0$. Meanwhile, there may still exist BPS states, i.e., forms annihilated by
\be
\Delta_{BPS} = D D^* - D^* D \,.
\ee
As in the free case, a massive BPS eigenstate $\Psi_\alpha$ of $H$ and $P$ is annihilated by an operator
\be
e^{i\theta_\alpha} D + e^{-i\theta_\alpha} D^* \,.
\ee

Let us end by generalizing what we have learned about $\cN=(1,1)$ theories to theories with $\cN=(2,2)$ supersymmetry. The target space $X$ of an $\cN = (2,2)$ sigma model is K\"ahler, and the free loop space $\cL X$ inherits this structure. Let us construct the supersymmetry algebra of a component $\cL X|_{\vec Q = \vec w}$ with non-trivial topological charge from linear forms, which involve the Dolbeault operators $\p$ and $\bp$.

Let us consider a Killing vector field $v$ on the K\"ahler ``manifold" $\cL X|_{\vec w}$. The vector $v$ preserves the complex structure, and is decomposable into holomorphic and anti-holomorphic Killing vectors $V$ and $\bar V$:
\be
v = V + \bar V, \quad
\{ \iota_V, \bp \} = \{ \iota_{\bar V}, \p \} = 0\,.
\ee
The vector fields may be written locally as $V = V^I \p_I$, $\bV = \bV^{\bar I} \p_{\bar I} = \overline{V^{I}} \p_{\bar I}$, where $I$ and $\bar I$ label the local holomorphic and anti-holomorphic coordinates $z^I$ and $\bar z^{\bar I}$ of $\cL X|_{\vec w}$. Since $d = \p + \bar \p$, it follows that
\be
\{ \iota_V, \p \} = \cL_{V},\quad \{ \iota_{\bar V}, \bp \} = \cL_{\bV} \,.
\ee
Since $V$ and $\bar V$ themselves are Killing vectors,
\be
\cL_{V}^* = -\cL_{V},\quad
\cL_{\bV}^* = -\cL_{\bV} \,.
\ee 

The $\cN = (2,2)$ supersymmetry operators on $\cL X|_{\vec w}$ can be constructed by using a flat connection
\be
A^U=U_{\bar I} d {\bar z}^{\bar I}
\ee
on $\cL X|_{\vec w}$, and the Killing vector $v_K=V + \bV$  inherited from the isometry of $S^1$.
\be
A^U \wedge = \iota_{U}^* \quad
(U_{\bar I} = G_{\bar I J} U^J)
\ee
for a vector field $U$ with only components with holomorphic indices, i.e., $U = U^I \p_I$. Here, $G_{I \bar J}=G_{\bar J I}$ are the compononets of the K\"ahler metric on $\cL X$. Given that the connection $A^U$ satisfies
\be\label{AU conditions}
\cL_{V} A^U = \cL_{\bar V} A^U =0, \quad
\p A^U = \bp A^U= 0 \,,
\ee
the pairing
\be
Z \equiv (-2i)\iota_\bV A^{U} = (-2i) \iota_{U} A^\bV 
\ee
is a constant complex number, while $\cL_{v_K} A^U =\cL_{v_K} A^\bU =0$. For $\cN=(2,2)$ non-linear sigma models into K\"ahler manifolds with holomorphic superpotentials, the Killing vectors and flat connections can be written explicitly, and the conditions \eq{AU conditions} can be shown to hold, following similar steps to the proof of the properties \eq{Au properties}. This is demonstrated in appendix \ref{ap:diff loop}.

Let us now define the differential operators 
\bea\label{D ops}
\cD &= \p + \iota_\bV + A^U \wedge \,, &
\cD^\dagger &= \p^\dagger + A^V \wedge + \iota_{\bU} \,, \\
\bar \cD &= \bp + \iota_V + A^\bU \wedge \,, &
\bar \cD^\dagger &= \bp^\dagger + A^\bV \wedge + \iota_{U} \,,
\eea
where we have modified equivariant Dolbeault operators \cite{Teleman,Lillywhite} by a connection. The dagger denotes the adjoint action on an operator $\cO$ such that
\be
\int \cO^\dagger \alpha \wedge * \bar\beta \equiv
\int \alpha \wedge * \overline{\cO \beta} \,.
\ee
Note that $\cO^* = \cO^\dagger$ for real operators, by which we mean operators such that for any real differential form $\beta$, $\cO \beta$ is also real. It is useful to note that $\p^\dagger = (\bp)^*$ and $\bp^\dagger = (\p)^*$. We then arrive at the commutation relations
\bea\label{D-comm}
&\{ \cD, \cD^\dagger \} =
\{ \bar \cD, \bar \cD^\dagger \} = H \,, \\
&\{  \cD, \bar \cD \} = 
- \{ \cD^\dagger, \bar \cD^\dagger \} = iP \equiv \cL_{v_K} \,, \\
&\cD^2 = (\bar \cD^\dagger)^2 = iZ/2 \,, \quad
\bar \cD^2 = (\cD^\dagger)^2 = -i\bar Z/2 \,,
\eea
while the anti-commutators  $\{ \cD, \bar \cD^\dagger \}$ and $\{ \bar \cD, \cD^\dagger \}$ vanish. The Hamiltonian $H$ is given by
\bea
2H=& \Delta + |v_K|^2 + |u|^2  + \{ d, A^{v_K} \wedge \}+\{ d, A^{v_K} \wedge \}^* \\
&+ \{ d, \iota_{u} \}+\{ d, \iota_u \}^*  \,,
\eea
where we have defined $u \equiv U+\bar U$. Details of the derivation of these relations are given in appendix \ref{ap:diff loop}.
Defining the operators
\bea
\bar \cQ_+ &= i^{1/2} \cD + i^{-1/2} \bar \cD^\dagger \,,&
\cQ_+ &= i^{1/2} \bar \cD + i^{-1/2} \cD^\dagger \,, \\
\bar \cQ_- &= i^{1/2} \cD^\dagger + i^{-1/2} \bar \cD \,,&
\cQ_- &= i^{1/2} \bar \cD^\dagger + i^{-1/2} \cD \,,
\eea
we find that they satisfy the supersymmetry algebra \eq{susy} with all other anti-commutators vanishing.%
~\footnote{We see that the bar used on the supersymmetry operators actually corresponds to taking the adjoint operation on the differential operators, rather than complex conjugation. Hence the notation $\bar \cQ_+$ and $\bar \cQ_-$ is misleading, although we have chosen to use it for sake of consistency with the literature.}

When $Z$ is non-zero, the operators $\cD$ or $\bar \cD$ do not define a cohomology, due to the last line of \eq{D-comm}. In particular, there are no states that are annihilated by the Hamiltonian $H$. There exists, however, a quadratic Hermitian operator that annihilates the BPS states given by
\be
\Delta_{BPS} = D_\zeta D_\zeta^\dagger -
D_\zeta^\dagger D_\zeta
\ee
where $2\zeta$ is the phase of $iZ$, i.e, $iZ = |Z| e^{2 i\zeta}$ while
\be
D_\varphi \equiv e^{-i\varphi} \cD + e^{i\varphi} \bar \cD \,,\quad
D_\varphi^\dagger = e^{i\varphi}\cD^\dagger + e^{-i\varphi} \bar \cD^\dagger \,.
\ee
Since $\Delta_{BPS}$ is Hermitian and
\be
\Delta_{BPS}^2 = 4(H^2 - P^2 - |Z|^2 ) \,,
\ee
the condition that a state is annihilated by $\Delta_{BPS}$ is equivalent to it having mass $|Z|$, as desired. Furthermore, following similar steps to the $\cN=(1,1)$ case, an eigenstate $\Psi_{h,p}$ of $H$ and $P$ that is BPS must be annihilated by the two Hermitian operators
\bea\label{ann}
e^{i\theta} D_\zeta + e^{-i\theta} D_\zeta^\dagger \,, ~~
e^{-i(\theta-{\pi \ov 2})} D_{(\zeta+{\pi \ov 2})} + e^{i(\theta-{\pi \ov 2})} D_{(\zeta+{\pi \ov 2})}^\dagger
\eea
with $e^{2 i \theta} = {(ip -|Z|)/h}$ and vice versa.

We have constructed the supersymmetry operators by modifying the equivariant differential operators on loop space by a flat connection. In topologically non-trivial sectors of the loop-space, in which the central charge $Z$ is non-zero, this connection has a non-trivial holonomy around the one-cycles given by the orbits of the Killing vector field $v_K$. From these operators, we can construct a quadratic operator $\Delta_{BPS}$ that annihilates the BPS states of the theory. It would be interesting to gain a better understanding of the differential geometry of loop space and in particular, the mathematical significance of the operator $\Delta_{BPS}$.

\appendix
\section{The BPS condition as a linear constraint}\label{ap:a}

In this section, we show that given an eigenstate $\Psi_{H=h,P=p}$ of $H$ and $P$, that the condition
\be
\Delta_{BPS} \Psi_{h,p} = 0
\ee
is equivalent to the condition that there exist real $a$ and $b$ such that
\be
\cQ(a,b) \equiv a \cQ_1 + b \cQ_2
\ee
annihilates $\Psi_{h,p}$.

Let us consider the action of $\cQ(a,b)^2$ on complex-valued differential forms. Using the definitions \eq{QD} and \eq{Ddef} we find:
\be
\int \cQ(a,b) \phi \wedge * \overline{\cQ(a,b) \phi} 
= \int \cQ(a,b)^2 \phi \wedge * \overline{\phi}
\,
\ee
Therefore $\cQ(a,b)^2$ is a positive semi-definite operator and
\be
\cQ(a,b)^2 \Psi_{h,p} = 0
\quad \Leftrightarrow \quad
\cQ(a,b) \Psi_{h,p} = 0 \,.
\label{Q2Q}
\ee

By the supersymmetry algebra, we can show that
\be
\cQ(a,b)^2 \Psi_{h,p} = \left[ (h-p) a^2 + 4 \pi Mwsab + (h+p) b^2 \right] \Psi_{h,p} \,.
\label{Q2}
\ee
Hence the existence of real $a, b$ for which
\be
\cQ(a,b)^2 \Psi_{h,p} = 0 \,,
\ee
is equivalent to the condition
\be
h^2 - p^2 - (2 \pi ws M)^2 \leq 0 \,.
\ee
The left hand side of this equation is precisely given by $4 \pi^2 \Delta_{BPS}^2$:
\be
\Delta_{BPS}^2 \leq 0\,.
\ee
Recall that the eigenstate $\Psi_{h,p}$ of $H$ and $P$ has to be an eigenstate of $\Delta_{BPS}^2$. Since $\Delta_{BPS}$ is Hermitian, we find that this is equivalent to
\be
\Delta_{BPS} \Psi_{h,p} = 0\,.
\ee
Combining this with the result \eq{Q2Q}, we arrive at our claim.

It is useful to note that defining the angle
\be
\theta = {\pi \ov 4} + \arg (a-ib) \,,
\ee
we find that
\be
e^{2i\theta} = i \left( {a^2 -b^2 -2iab \ov a^2 + b^2} \right) = {2\pi wsM + ip \ov h}
\ee
and the operator $\cQ(a,b)$ can be written as
\be
\cQ(a,b)= \sqrt{a^2 +b^2} (e^{i\theta} D + e^{-i\theta}D^*) \,.
\ee

\section{The differential geometry of loop space}\label{ap:diff loop}

In this appendix, we present some relevant facts about the differential geometry of the loop space $\cL X$. In particular, we write explicit expressions for vector fields and one-forms in loop space that are relevant for defining the supersymmetry operators, and verify that they have desirable properties. In particular, we show some identities necessary for arriving at the commutation relations \eq{SUSY11} and \eq{D-comm} of differential operators defined in $\cL X$.

Let us denote the real bosonic fields mapping into the target space $X$ as $\phi^\mu(s)$, where $s$ is the coordinate on the unit circle. The pair $(\mu,s)$ constitutes a coordinate index in the infinite dimensional manifold $\cL X$, while the values $\phi^\mu(s)$ are real coordinates on this space. A vector field $V$, thus, has the component expression
\be
V = \int_{S^1} ds V^{\mu,s}  \partial_{\phi^\mu(s)} \,,
\ee
while a one-form $A$ can be written as
\be
A = \int_{S^1} ds A_{\mu,s} \delta {\phi^\mu(s)} \,.
\ee
The sum over the repeated indices are assumed. The coordinate dependence of the components $V^{\mu,s}$ and $A_{\mu,s}$ are implicit. The differential operator $d$ acts on the one-form $A$ as
\be
dA = \int_{S^1} ds \partial_{\phi^\nu(s)} A_{\mu,s} \delta {\phi^\nu(s)} \wedge \delta {\phi^\mu(s)} \,,
\ee
and similarly on differential forms in general. The metric on $\cL X$ is defined by a pull-back from the target space $X$:
\be
V \cdot W (\phi(s)) = \int_{S^1} ds V^{\mu,s} W^{\nu,s} G_{\mu\nu} (\phi(s)) \,.
\ee
In components, the metric can be written as
\be
G_{(\mu,s)(\nu,s')} (\phi(s))= G_{\mu\nu} (\phi(s))\delta(s-s') \,.
\ee

Now the components of the Killing vector $v_K$, defined by the pull-back of the action of the translation along the circle to the loop space, can be explicitly written as
\be
v_K = \int_{S^1} ds {d \phi^\mu(s) \ov ds} \partial_{\phi^\mu (s) } \,.
\ee
The one-form $A^u$ used in defining the supersymmetry algebra is given by
\be
A^u = \int_{S^1} ds \partial_\mu \cV (\phi (s) ) \delta \phi^\mu(s) \,,
\ee
where $\cV$ is the $\cN=(1,1)$ superpotential. The connection $A^u$ is closed:
\be
d A^u = \int_{S^1} ds \partial_{\nu} \partial_\mu \cV (\phi (s) ) \delta \phi^\nu(s) \wedge \delta \phi^\mu(s)  = 0 \,.
\ee
Also, by definition,
\be
\iota_{v_K} A^u = \int_{S^1} ds {d \phi^\mu(s) \ov ds} \partial_\mu \cV (\phi (s) ) = \Delta \cV
\ee
is a constant determined by the topological winding number of the component $\cL X|_{\vec Q = \vec w}$ of the loop space. It follows that
\be
\cL_{v_K} A^u = \{ d, \iota_{v_K} \} A^u = 0 \,.
\ee
We have thus proven the properties \eq{Au properties} of $A^u$.

Let us now consider the case of a sigma model with $\cN=(2,2)$ supersymmetry. The target space $X$, and correspondingly, the loop space $\cL X$ are now K\"ahler. Now $\phi^i (s)$ denote the complex bosonic component of the (twisted) chiral fields, while $i, j, \cdots$ and $\bar i, \bar j, \cdots$ are used to index the holomorphic and anti-holomorphic coordinates, respectively. Now the pair of coordinates $(i,s)$ makes up the holomorphic indices on $\cL X$, while $(\bar i, s)$ makes up the anti-holomorphic indices. The Killing vector $v_K = V +\bar V$ splits into a sum of a holomorphic and anti-holomorphic Killing vector. $V$ and $\bar V$ can be explicitly written as
\be
V = \int_{S^1} ds {d \phi^i(s) \ov ds} \partial_{\phi^i (s) }, \quad
\bar V = \int_{S^1} ds {d \bar\phi^{\bar i}(s) \ov ds} \partial_{\bar\phi^{\bar i} (s) } \,.
\ee
The one-form $A^U$ in equation \eq{D ops} is given by
\be
A^{\bar U} = \int_{S^1} ds \partial_i \tW (\phi (s) ) \delta \phi^i(s) \,,
\ee
while $A^{ U} = \overline{A^{\bar U}}$ for the (twisted) superpotential $\tW$. Recall that $\tW$ is a holomorphic function of the complex fields $\phi^i$. Using these definitions, the relations \eq{AU conditions} can be proven by following the steps taken in the $\cN=(1,1)$ case. First of all,
\bea\label{flat}
\bp A^{\bar U} &= \int_{S^1} ds \, \bp_{\bar j} \p_i \tW \,
\delta \bar\phi^{\bar j} (s) \wedge \delta \phi^i (s) = 0 \,, \\
\p A^{\bar U} &= \int_{S^1} ds \, \p_{j} \p_i \tW \,
\delta \phi^{ j} (s) \wedge \delta \phi^i (s) = 0 \,.
\eea
Meanwhile, $\iota_{\bar V} A^{\bar U} = 0$ while
\bea
\iota_{V} A^{\bar U} = \int_{S^1} ds  {d \phi^i(s) \ov ds }\p_i \tW(\phi(s))  = \Delta \tW \,,
\eea
which is constant in each component of $\cL X$. Thus \eq{AU conditions} follows.

Some additional identities need to be shown in order to arrive at the commutation relations \eq{D-comm}. To show the vanishing of the commutators $\{ \cD, \bar \cD^\dagger \}$ and $\{ \bar \cD, \cD^\dagger \}$, the following must hold:
\be\label{vanishing relations}
\{\p,A^{\bar V} \wedge \} = (\p A^{\bar V}) \wedge =0 \,, \quad
\{ \p, \iota_U \} + \{ \p, \iota_U \}^* = 0 \,.
\ee
The first identity follows from the fact that the target manifold is K\"ahler, that is:
\bea
\p A^{\bar V} &= \int_{S^1} ds \p_k
\left( G_{i \bar j} {d \bar \phi^{\bar j} (s) \ov ds} \right)
\delta \phi^{k}(s) \wedge \delta \phi^{i}(s) \\
&= \int_{S^1} ds \left( \p_k \p_i \bp_{\bar j}K \right)
{d \bar \phi^{\bar j} (s) \ov ds}
\delta \phi^{k}(s) \wedge \delta \phi^{i}(s)  = 0\,.
\eea
Here the fact that the K\"ahler metric is given by $G_{i \bar j} = \p_i \bp_{\bar j}K$ for the K\"ahler potential $K$ has been used.

The condition $\bp A^{U} =0$ is enough to ensure that the second equation of \eq{vanishing relations} holds. Let us denote all coordinate indices of $\cL X$ using $M,N, \cdots$ while using the letters $I,J, K, L$ and $\bar I, \bar J, \bar K, \bar L$ to denote holomorphic and anti-holomorphic indices, respectively. These indices have a continuous range, and the sum/integral over the repeated indices are assumed in the equations to follow. We also denote the holomorphic (anti-holomorphic) local coordinates by $z^I$ ($\bar z ^{\bar I}$), respectively. The action of $\{ \p,\iota_U\}$ on a differential form $T$ with components $T_{M_1 M_2 \cdots}$ is then given by
\bea
&\left[ \{ \p,\iota_U\} T \right] _{M_1 M_2 \cdots }  = U^I \p_I T_{M_1 M_2 \cdots} \\
&+ \sum_{M_r \text{: holomorphic index}} (\p_{M_r} U^I) T_{M_1 \cdots M_{r-1} I M_{r+1} \cdots} \,.
\eea
Recall that the components of $A^U = U_{\bar I} d\bar z^{\bar I}$ are related to the components of the vector $U = U^I \p_I$ by $U_{\bar I} = G_{\bar I J} U^J$ while $\p_{J} U_{\bar I} = 0 $ for all pairs of $\bar I$ and $J$. Using this fact and the fact that the only non-zero Levi-Civit\'a connections of a K\"ahler manifold are given by
\be\label{Kahler connection}
\Gamma^I_{JK} = - G_{K \bar L} \p_J G^{I\bar L} \,, \quad
\Gamma^{\bar I}_{\bar J \bar K} = - G_{\bar K L} \bp_{\bar J} G^{\bar I L} \,,
\ee
we arrive at
\be\label{rel 1}
 \{ \p,\iota_U\} T  = U^M \nabla_M T \,.
\ee
It can also be shown using the same set of facts that
\be\label{rel 2}
\nabla_M U^M =0 \,.
\ee
Equations \eq{rel 1} and \eq{rel 2} imply the equality $\{ \p,\iota_U\} = - \{ \p,\iota_U\}^*$, since
\bea
&\int (U^M \nabla_M \alpha) \wedge * \beta = \\
&-\int (\nabla_M U^M)  (\alpha \wedge * \beta)
-\int \alpha \wedge * (U^M \nabla_M \beta) \,.
\eea

To show that the commutation relation
\be
\{ \cD, \cD^\dagger \} =
\{ \bar \cD, \bar \cD^\dagger \} = H
\ee
holds, the following equalities must also be verified:
\be\label{for H}
\{ \p , A^V \wedge \} = \{ \bp, A^{\bar V} \wedge \} \,, \quad
\{ \bp , \iota_U \} = \{ \bp , \iota_U \}^* \,.
\ee
The first of these equations follows from the fact that $v_K$ is a Killing vector, i.e., $\nabla_M ({v_K})_N +\nabla_N ({v_K})_M =0$. This implies that
\be
\p A^V = \p_I V_{\bar J} dz^I \wedge d\bar z^{\bar J}
= \p_{\bar J} {\bar V}_{I} d\bar z^{\bar J} \wedge d z^{I} = \bp A^{\bar V} \,.
\ee
Meanwhile, the action of the operator $\{ \bp, \iota_U \}$ on a differential form $T$ is given by
\bea\label{anti Lie}
&\left[ \{ \bp,\iota_U\} T \right] _{M_1 M_2 \cdots }  = \\
& \sum_{M_r \text{: anti-holomorphic index}} (\bp_{M_r} U^I) T_{M_1 \cdots M_{r-1} I M_{r+1} \cdots} \,.
\eea
Using the flatness of $A^U$, this equation can be rewritten as
\bea
\left[ \{ \bp,\iota_U\} T \right] _{M_1 M_2 \cdots }  =
 \sum_{r} ( \nabla^M U_{M_r}) T_{M_1 \cdots M_{r-1} M M_{r+1} \cdots} \,.
\eea
From this equation and the definition of the action of $*$ on operators, it follows that
\bea\label{anti Lie dual}
\left[ \{ \bp,\iota_U\}^* T \right] _{M_1 M_2 \cdots }  =
 \sum_{r} ( \nabla_{M_r} U^M) T_{M_1 \cdots M_{r-1} M M_{r+1} \cdots} \,.
\eea
Meanwhile, the flatness of $A^U$ along with equation \eq{Kahler connection} is enough to arrive at
\be\label{nabla U}
 \nabla_{M_r} U^M =
 \begin{cases}
 \bp_{\bar J} U^I & \text{when $M_r = \bar J$ and $M = I$} \\
 0 & \text{otherwise} \,,
 \end{cases}
\ee
that is, $\nabla_{M_r} U^M$ vanishes unless $M_r$ is an anti-holomorphic index and $M$ is a holomorphic index. Comparing equations \eq{anti Lie}, \eq{anti Lie dual} and \eq{nabla U}, we arrive at the second identity of equation \eq{for H}.
\vspace*{0.5in}

\begin{acknowledgments}
I would like to thank A. Abanov, C. Closset, C. Herzog, K. Jensen, A. Kapustin, J. Lee, P. Longhi, C. Meneghelli, G. W. Moore, C. Y. Park, L. Rastelli, E. Witten and especially the referee of this paper for illuminating discussions and/or useful comments that have gone into significantly improving the draft. I would also like to thank the New High Energy Theory Center at Rutgers University and the School of Physics at KIAS for their hospitality during the completion of this work. This work is supported in part by DOE grant DE-FG02-92ER-40697.
\end{acknowledgments}

\bibliography{Top}

\providecommand{\href}[2]{#2}\begingroup\raggedright\begin{thebibliography}{10}

\bibitem{Skyrme}
T.~Skyrme, ``{A Nonlinear field theory},''
\href{http://dx.doi.org/10.1098/rspa.1961.0018}{{\em Proc.Roy.Soc.Lond.}
  {\bfseries A260} (1961) 127--138}.

\bibitem{FinkelsteinRubinstein}
D.~Finkelstein and J.~Rubinstein, ``{Connection between spin, statistics, and
  kinks},''
\href{http://dx.doi.org/10.1063/1.1664510}{{\em J.Math.Phys.} {\bfseries 9}
  (1968) 1762--1779}.

\bibitem{Faddeev}
L.~Faddeev, ``{Some Comments on the Many Dimensional Solitons},''
\href{http://dx.doi.org/10.1007/BF00398483}{{\em Lett.Math.Phys.} {\bfseries 1}
  (1976) 289}.

\bibitem{Coleman}
S.~R. Coleman, ``{The Quantum Sine-Gordon Equation as the Massive Thirring
  Model},''
\href{http://dx.doi.org/10.1103/PhysRevD.11.2088}{{\em Phys.Rev.} {\bfseries
  D11} (1975) 2088}.

\bibitem{JackiwRebbi}
R.~Jackiw and C.~Rebbi, ``{Spin from Isospin in a Gauge Theory},''
\href{http://dx.doi.org/10.1103/PhysRevLett.36.1116}{{\em Phys.Rev.Lett.}
  {\bfseries 36} (1976) 1116}.

\bibitem{HasenfratztHooft}
P.~Hasenfratz and G.~'t~Hooft, ``{A Fermion-Boson Puzzle in a Gauge Theory},''
\href{http://dx.doi.org/10.1103/PhysRevLett.36.1119}{{\em Phys.Rev.Lett.}
  {\bfseries 36} (1976) 1119}.

\bibitem{GoldstoneWilczek}
J.~Goldstone and F.~Wilczek, ``{Fractional Quantum Numbers on Solitons},''
\href{http://dx.doi.org/10.1103/PhysRevLett.47.986}{{\em Phys.Rev.Lett.}
  {\bfseries 47} (1981) 986--989}.

\bibitem{WittenBaryons}
E.~Witten, ``{Current Algebra, Baryons, and Quark Confinement},''
\href{http://dx.doi.org/10.1016/0550-3213(83)90064-0}{{\em Nucl.Phys.}
  {\bfseries B223} (1983) 433--444}.

\bibitem{WittenOlive}
E.~Witten and D.~I. Olive, ``{Supersymmetry Algebras That Include Topological
  Charges},''
\href{http://dx.doi.org/10.1016/0370-2693(78)90357-X}{{\em Phys.Lett.}
  {\bfseries B78} (1978) 97}.

\bibitem{HLS}
X.-r. Hou, A.~Losev, and M.~A. Shifman, ``{BPS saturated solitons in N=2
  two-dimensional theories on R x S: domain walls in theories with compactified
  dimensions},'' \href{http://dx.doi.org/10.1103/PhysRevD.61.085005}{{\em
  Phys.Rev.} {\bfseries D61} (2000) 085005},
\href{http://arxiv.org/abs/hep-th/9910071}{{\ttfamily arXiv:hep-th/9910071
  [hep-th]}}.

\bibitem{BSV}
D.~Binosi, M.~A. Shifman, and T.~ter Veldhuis, ``{Leaving the BPS bound:
  Tunneling of classically saturated solitons},''
  \href{http://dx.doi.org/10.1103/PhysRevD.63.025006}{{\em Phys.Rev.}
  {\bfseries D63} (2001) 025006},
\href{http://arxiv.org/abs/hep-th/0006026}{{\ttfamily arXiv:hep-th/0006026
  [hep-th]}}.

\bibitem{FMVW}
P.~Fendley, S.~Mathur, C.~Vafa, and N.~Warner, ``{Integrable Deformations and
  Scattering Matrices for the $N=2$ Supersymmetric Discrete Series},''
\href{http://dx.doi.org/10.1016/0370-2693(90)90848-Z}{{\em Phys.Lett.}
  {\bfseries B243} (1990) 257--264}.

\bibitem{WittenMorse}
E.~Witten, ``{Supersymmetry and Morse theory},''
{\em J.Diff.Geom.} {\bfseries 17} (1982) 661--692.

\bibitem{MooreFlorida}
G.~W. Moore, ``{Three Lectures on Fukaya-Seidel Categories and Web-Based
  Formalism}.''
\newblock \url{http://www.physics.rutgers.edu/~gmoore/FloridaLectures2.pdf}.

\bibitem{GMWweb}
D.~Gaiotto, G.~W. Moore, and E.~Witten, ``{An Introduction To The Web-Based
  Formalism},''
\href{http://arxiv.org/abs/1506.04086}{{\ttfamily arXiv:1506.04086 [hep-th]}}.

\bibitem{GMW}
D.~Gaiotto, G.~W. Moore, and E.~Witten, ``{Algebra of the Infrared: String
  Field Theoretic Structures in Massive ${\cal N}=(2,2)$ Field Theory In Two
  Dimensions},''
\href{http://arxiv.org/abs/1506.04087}{{\ttfamily arXiv:1506.04087 [hep-th]}}.

\bibitem{CVClassification}
S.~Cecotti and C.~Vafa, ``{On classification of N=2 supersymmetric theories},''
  \href{http://dx.doi.org/10.1007/BF02096804}{{\em Commun.Math.Phys.}
  {\bfseries 158} (1993) 569--644},
\href{http://arxiv.org/abs/hep-th/9211097}{{\ttfamily arXiv:hep-th/9211097
  [hep-th]}}.

\bibitem{Teleman}
C.~Teleman, ``The quantization conjecture revisited,''
  \href{http://dx.doi.org/10.2307/2661378}{{\em Ann. of Math. (2)} {\bfseries
  152} no.~1, (2000) 1--43}. \url{http://dx.doi.org/10.2307/2661378}.

\bibitem{Lillywhite}
S.~Lillywhite, ``Formality in an equivariant setting,''
  \href{http://dx.doi.org/10.1090/S0002-9947-03-03265-3}{{\em Trans. Amer.
  Math. Soc.} {\bfseries 355} no.~7, (2003) 2771--2793}.
  \url{http://dx.doi.org/10.1090/S0002-9947-03-03265-3}.

\bibitem{CFIV}
S.~Cecotti, P.~Fendley, K.~A. Intriligator, and C.~Vafa, ``{A New
  supersymmetric index},''
  \href{http://dx.doi.org/10.1016/0550-3213(92)90572-S}{{\em Nucl.Phys.}
  {\bfseries B386} (1992) 405--452},
\href{http://arxiv.org/abs/hep-th/9204102}{{\ttfamily arXiv:hep-th/9204102
  [hep-th]}}.

\bibitem{CVIsing}
S.~Cecotti and C.~Vafa, ``{Ising model and N=2 supersymmetric theories},''
  \href{http://dx.doi.org/10.1007/BF02098023}{{\em Commun.Math.Phys.}
  {\bfseries 157} (1993) 139--178},
\href{http://arxiv.org/abs/hep-th/9209085}{{\ttfamily arXiv:hep-th/9209085
  [hep-th]}}.

\bibitem{Bershadsky:1993ta}
M.~Bershadsky, S.~Cecotti, H.~Ooguri, and C.~Vafa, ``{Holomorphic anomalies in
  topological field theories},''
  \href{http://dx.doi.org/10.1016/0550-3213(93)90548-4}{{\em Nucl.Phys.}
  {\bfseries B405} (1993) 279--304},
\href{http://arxiv.org/abs/hep-th/9302103}{{\ttfamily arXiv:hep-th/9302103
  [hep-th]}}.

\bibitem{SW1}
N.~Seiberg and E.~Witten, ``{Electric - magnetic duality, monopole
  condensation, and confinement in N=2 supersymmetric Yang-Mills theory},''
  \href{http://dx.doi.org/10.1016/0550-3213(94)90124-4}{{\em Nucl.Phys.}
  {\bfseries B426} (1994) 19--52},
\href{http://arxiv.org/abs/hep-th/9407087}{{\ttfamily arXiv:hep-th/9407087
  [hep-th]}}.

\bibitem{SW2}
N.~Seiberg and E.~Witten, ``{Monopoles, duality and chiral symmetry breaking in
  N=2 supersymmetric QCD},''
  \href{http://dx.doi.org/10.1016/0550-3213(94)90214-3}{{\em Nucl.Phys.}
  {\bfseries B431} (1994) 484--550},
\href{http://arxiv.org/abs/hep-th/9408099}{{\ttfamily arXiv:hep-th/9408099
  [hep-th]}}.

\bibitem{Witten:1997sc}
E.~Witten, ``{Solutions of four-dimensional field theories via M theory},''
  \href{http://dx.doi.org/10.1016/S0550-3213(97)00416-1}{{\em Nucl.Phys.}
  {\bfseries B500} (1997) 3--42},
\href{http://arxiv.org/abs/hep-th/9703166}{{\ttfamily arXiv:hep-th/9703166
  [hep-th]}}.

\bibitem{Gaiotto:2009we}
D.~Gaiotto, ``{N=2 dualities},''
  \href{http://dx.doi.org/10.1007/JHEP08(2012)034}{{\em JHEP} {\bfseries 1208}
  (2012) 034},
\href{http://arxiv.org/abs/0904.2715}{{\ttfamily arXiv:0904.2715 [hep-th]}}.

\bibitem{Alday:2009aq}
L.~F. Alday, D.~Gaiotto, and Y.~Tachikawa, ``{Liouville Correlation Functions
  from Four-dimensional Gauge Theories},''
  \href{http://dx.doi.org/10.1007/s11005-010-0369-5}{{\em Lett.Math.Phys.}
  {\bfseries 91} (2010) 167--197},
\href{http://arxiv.org/abs/0906.3219}{{\ttfamily arXiv:0906.3219 [hep-th]}}.

\bibitem{GGS}
D.~Gaiotto, S.~Gukov, and N.~Seiberg, ``{Surface Defects and Resolvents},''
  \href{http://dx.doi.org/10.1007/JHEP09(2013)070}{{\em JHEP} {\bfseries 1309}
  (2013) 070},
\href{http://arxiv.org/abs/1307.2578}{{\ttfamily arXiv:1307.2578}}.

\bibitem{WittenGLSM}
E.~Witten, ``{Phases of N=2 theories in two-dimensions},''
  \href{http://dx.doi.org/10.1016/0550-3213(93)90033-L}{{\em Nucl.Phys.}
  {\bfseries B403} (1993) 159--222},
\href{http://arxiv.org/abs/hep-th/9301042}{{\ttfamily arXiv:hep-th/9301042
  [hep-th]}}.

\bibitem{HoriVafa}
K.~Hori and C.~Vafa, ``{Mirror symmetry},''
\href{http://arxiv.org/abs/hep-th/0002222}{{\ttfamily arXiv:hep-th/0002222
  [hep-th]}}.

\bibitem{RocekVerlinde}
M.~Rocek and E.~P. Verlinde, ``{Duality, quotients, and currents},''
  \href{http://dx.doi.org/10.1016/0550-3213(92)90269-H}{{\em Nucl.Phys.}
  {\bfseries B373} (1992) 630--646},
\href{http://arxiv.org/abs/hep-th/9110053}{{\ttfamily arXiv:hep-th/9110053
  [hep-th]}}.

\bibitem{dGIT}
J.~de~Azcarraga, J.~P. Gauntlett, J.~Izquierdo, and P.~Townsend, ``{Topological
  Extensions of the Supersymmetry Algebra for Extended Objects},''
\href{http://dx.doi.org/10.1103/PhysRevLett.63.2443}{{\em Phys.Rev.Lett.}
  {\bfseries 63} (1989) 2443}.

\bibitem{Gukov:2006jk}
S.~Gukov and E.~Witten, ``{Gauge Theory, Ramification, And The Geometric
  Langlands Program},''
\href{http://arxiv.org/abs/hep-th/0612073}{{\ttfamily arXiv:hep-th/0612073
  [hep-th]}}.

\bibitem{Gaiotto:2009fs}
D.~Gaiotto, ``{Surface Operators in N = 2 4d Gauge Theories},''
  \href{http://dx.doi.org/10.1007/JHEP11(2012)090}{{\em JHEP} {\bfseries 1211}
  (2012) 090},
\href{http://arxiv.org/abs/0911.1316}{{\ttfamily arXiv:0911.1316 [hep-th]}}.

\bibitem{Gaiotto:2010be}
D.~Gaiotto, G.~W. Moore, and A.~Neitzke, ``{Framed BPS States},''
  \href{http://dx.doi.org/10.4310/ATMP.2013.v17.n2.a1}{{\em
  Adv.Theor.Math.Phys.} {\bfseries 17} (2013) 241--397},
\href{http://arxiv.org/abs/1006.0146}{{\ttfamily arXiv:1006.0146 [hep-th]}}.

\bibitem{GMN}
D.~Gaiotto, G.~W. Moore, and A.~Neitzke, ``{Wall-Crossing in Coupled 2d-4d
  Systems},'' \href{http://dx.doi.org/10.1007/JHEP12(2012)082}{{\em JHEP}
  {\bfseries 1212} (2012) 082},
\href{http://arxiv.org/abs/1103.2598}{{\ttfamily arXiv:1103.2598 [hep-th]}}.

\bibitem{Polchinski1}
J.~Polchinski, {\em String Theory}, vol.~1.
\newblock Cambridge University Press, 1998.
\newblock \url{http://dx.doi.org/10.1017/CBO9780511816079}.

\bibitem{Polchinski2}
J.~Polchinski, {\em String Theory}, vol.~2.
\newblock Cambridge University Press, 1998.
\newblock \url{http://dx.doi.org/10.1017/CBO9780511618123}.

\bibitem{GriffithsHarris}
P.~Griffiths and J.~Harris, \href{http://dx.doi.org/10.1002/9781118032527}{{\em
  Principles of algebraic geometry}}.
\newblock Wiley Classics Library. John Wiley \& Sons, Inc., New York, 1994.
\newblock \url{http://dx.doi.org/10.1002/9781118032527}.
\newblock Reprint of the 1978 original.

\end{thebibliography}\endgroup

\end{document}